\documentstyle[epsfig,aps]{revtex}
\newcommand{\bba}{\begin{eqnarray}}
\newcommand{\eea}{\end{eqnarray}}
\newcommand{\bb}{\begin{equation}}
\newcommand{\ee}{\end{equation}}
\newcommand{\bban}{\begin{eqnarray*}}
\newcommand{\eean}{\end{eqnarray*}}

\newcommand{\barI}{I{\!\!\!}^{\_}}
\def\a{\alpha}
\def\b{\beta}
\def\g{\gamma}

\def\d{\delta}

\def\f{\phi}

\def\l{\lambda}
\def\m{\mu}

\def\r{\rho}
\def\s{\sigma}

\def\x{\xi}

\def\G{\Gamma}

\def\O{\Omega}

\title{Cosmological Perturbations in Brane Worlds:\\ 
 Brane Bending and Anisotropic Stresses}
\author{Miquel Dorca$^1$\footnote{email: dorca@het.brown.edu} and 
Carsten van de Bruck$^{2}$\footnote{email: C.VanDeBruck@damtp.cam.ac.uk},
\\
$^{1}${\small Department of Physics, Brown University, 182 Hope
Street, Providence, RI 02912, USA}\\
$^{2}${\small DAMTP, Center for Mathematical Sciences, Wilberforce
       Road, CB3 0WA, Cambridge, United Kingdom}\\
\vspace{0.25cm}
}
\begin{document}
\noindent
\maketitle
\begin{abstract}
Using a metric--based formalism to treat cosmological perturbations, 
we discuss the connection between anisotropic stress on the brane and
brane bending. First we discuss gauge--transformations, and draw our 
attention to gauges, in which the brane--positions remain
unperturbed. We provide a unique gauge, where perturbations both on 
the brane and in the bulk can be treated with generality.
For vanishing anisotropic stresses on the brane, this gauge 
reduces to the generalized longitudinal gauge. We further comment on the 
gravitational interaction between the branes and the bulk. 

\end{abstract}

\section{Introduction}
One outstanding problem in brane world cosmology is to develop 
a better understanding on the evolution of perturbations. 
To gain insights is of prime importance. In fact, in order to make 
predictions regarding the primordial spectrum of perturbations or the
anisotropies in the cosmic microwave background, the creation and evolution of 
perturbations have to be understood. Only then we can hope to test the 
theory with future cosmological experiments. On the other hand, the 
question of stability of the global brane world space--time has to be 
addressed, because some (or most) solutions of Einstein equations
could be in principle unstable. To attack these issues, the use of a 
full five--dimensional description is necessary. 

Some work has already addressed fluctuations in brane world theories. 
Namely, creation of perturbations on the brane was considered in 
[\ref{wands}]--[\ref{soda2}], where specific assumptions were made from the 
beginning. Other groups have developed a formalism to treat perturbations 
in brane worlds for rather general situations, see [\ref{car00}] 
(hereafter Paper I) and [\ref{per11}]--[\ref{per6}]. 
In Paper I, a metric--based formalism, which represents 
a straightforward extension of the usual four--dimensional approach 
[\ref{Brandenberger}], was developed for a brane world theory with 
two branes. Also, some of the evolution
equations on the brane were found. However, two assumptions where
made, namely: i) it was required that the position of the branes 
was not affected by a first order perturbation, and 
ii) a generalized longitudinal gauge (GLG) was used in the bulk. 
A result of these assumptions was that the anisotropic stresses 
on the branes had necessarily to vanish. 
Although, this particular scenario would be adequate to study 
scalar fields or ordinary matter without anisotropic stresses, 
it does not represent the most general case.
The aim of this paper is precisely to fill this gap by 
considering the most general case of having anisotropic stresses on the 
branes and by allowing the brane positions to be perturbed. 

We will consider only the case of single extra dimension, which is
assumed to be compactified on a circle $S_1$ with a $Z_2$--symmetry. 
The (unperturbed) branes will be located at 
the fix points of the $Z_2$--symmetry, which, in our case are
taken to be $y_1=0$ and $y_2=R$. 

The paper is organized as follows: in the next section we will discuss 
gauge transformations. After briefly reviewing the construction 
of gauge invariant variables, we will show that the GLG is not compatible 
with unperturbed brane positions in the most general case. 
We will then discuss a novel gauge, which reduces to the GLG for 
vanishing anisotropic stresses on the branes. 
We will also discuss the Randall--Sundrum 
gauge in the context of cosmological perturbations. In Section III,
we will derive the perturbed Einstein equations for the gauge introduced 
in Section II and we will derive the junction conditions. We will also 
briefly discuss the gravitational coupling between the branes and the
bulk in Section IV. Finally, our conclusions can be found in Section V. 

\section{Gauge transformations and gauge-invariant variables}
In this section we begin with a brief review of the gauge-invariant 
formalism for metric perturbations in brane--world models (see Paper
I). Then we will discuss gauge--transformations and the inclusion of 
the anisotropic stresses. Throughout 
this paper, we will consider the case for five--dimensional brane worlds
only. The additional spatial coordinate is denoted by $y^5=y$. For 
the purpose of this section, all we need to specify is that the 
branes are stretched across the usual four-dimensional space-time and 
that they are located at specific points along the additional dimension. 
We will be more precise below.

The most general higher-dimensional metric consistent with the
maximally symmetric three-dimensional spatial manifold is given 
by (here the indices $a$ and $b$ are either 0 or 5 and $y^0 = t$)
\begin{equation}
ds^2 = a^2\left\{\gamma_{ab}dy^ady^b-\Omega_{ij}dx^idx^j\right\},
\end{equation}                   
where the scale factor $a$ and the metric $\g_{ab}$ are functions of
the coordinates $y^a$ only and $\O_{ij}$
is the metric of the three--dimensional maximally symmetric space, 
given by
\bb\label{Omega}
\Omega_{ij}=\delta _{ij}\left[1+{k\over 4}x^lx^m\d_{lm}\right]^{-2}\; ,
\ee
with $k=0,1,-1$ for flat, closed or hyperbolic 3--geometries,
respectively. 
Given this structure of the background, we are able to 
classify metric perturbations by their three-dimensional tensor
properties as in the four-dimensional case [\ref{Brandenberger}]. 
The perturbed metric can be generically written in the form
\begin{equation}\label{permetric1}
ds^2 =  a^2\left\{\gamma_{ac}\left( \delta^{c}_{b} +
        2\phi^{c}_{b}\right) dy^a dy^b 
       -\left[ \left(1 - 2\psi \right) \Omega_{ij} + 2E_{|ij} +
       2F_{(i|j)} + h_{ij}\right] dx^i dx^j-2 W_{ai}dy^a dx^i\right\}\; ,
\end{equation}
where  $F_i$ and $h_{ij}$ have a vanishing
divergence and $h_{ij}$ is traceless. 
Also, the three--vectors $W_{ai}$ can be split as follows, 
\bb
W_{ai} = B_{a|i} + S_{ai}\; ,
\ee
where ${{S_a}^i}_{|i}=0$. 

An infinitesimal coordinate transformation can be written as
\begin{eqnarray}\label{gaugetransf}
x^{\a} \rightarrow \tilde{x^\a} = x^{\a} + \xi^{\a} 
\end{eqnarray}
with (infinitesimal) parameters $\x^\a$. We split these 
parameters as $\x^\a  = (\x^a,\x^i)$, where 
$\xi ^i=\eta^i+\xi^{|i}$ being $\xi$ a scalar function
and $\eta^i$ a divergenceless
three--dimensional vector, i.e. $\eta^i_{~|i}=0$.
We use the convention that indices of type $a$ ($i$) are
lowered, raised and contracted using the metric $\g_{ab}$ ($\O_{ij}$).
Furthermore, we take the vertical bar to denote the covariant
derivative with respect to $\g_{ab}$ or $\O_{ij}$ depending on the
index type. As shown in [\ref{car00}], the transformation 
of the scalar sector is found to be
\begin{eqnarray}\label{gaugetransfg}
\delta \phi_{ab} &=& -\xi_{(a|b)}-H^c\x_c\g_{ab},\label{gaugetransfg1}\\
\delta \psi &=& H^a \xi_a,\label{gaugetransfg2}\\
\delta E &=& -\xi,\label{gaugetransfg3}\\
\delta B_a &=& \xi _{a} - \xi _{|a}\label{gaugetransfg4}.
\end{eqnarray}
where we have introduced the generalized Hubble parameters
\bb
 H_c=\frac{a_{|c}}{a}\; .
\ee
The vector perturbations in the metric (\ref{permetric1}) change according to
\begin{eqnarray}
\delta F_i &=& -\eta_i, \\
\delta S_{ai} &=& -\eta_{i|a}\; .
\end{eqnarray}
The tensor perturbation $h_{ij}$ is, of course, invariant under 
the first order gauge transformation. From these transformation rules one can easily 
construct gauge invariant variables for scalars and vectors, which 
we repeat here (see [\ref{car00}] for further details):
\begin{center}
\underline{{\it Scalar variables}}
\end{center}
\begin{eqnarray}\label{gaugeinvsca}
\qquad\quad\Phi_{ab} &=& \phi_{ab} + H^c(B_c-E_{|c})\g_{ab}+
                     (B_{(a}-E_{|(a})_{|b)}\\
\qquad\quad\Psi &=& \psi - H^c \left(B_c - E_{|c}\right).
\end{eqnarray}
\begin{center}
\underline{{\it Vector variables}}
\end{center}
\begin{eqnarray}\label{gaugeinvvec}
\qquad\quad {\cal F}_{ai} = S_{ai} - F_{i|a}.
\end{eqnarray}

In this paper we shall consider scalar perturbations only, for which 
the perturbed five-dimensional metric (\ref{permetric1})
reduces to
\bba
ds^2 &=& a^2\left\{ b^2\left[(1+2\phi)dt^2 -2Wdtdy 
       - (1-2\Gamma) dy^2 \right]\right.\nonumber\\
     &&\qquad\left. - \left[ \Omega _{ij}(1 - 2\psi) + 2 E_{|ij} 
       \right] dx^i dx^j-2B_{0|i}dt dx^i-2B_{5|i}dydx^i\right\}
  \label{met5}\,
\eea
where we have defined
\bb
 \f = \f_0^0\; ,\qquad \G = -\f_5^5\; ,\qquad W=2\f_0^5=-2\f_5^0\; .
\ee
It is convenient to use  a conformal gauge 
for the metric $\gamma_{ab}$, i.e.,
\bb
 (\g_{ab})=b^2\text{diag}(1,-1).
\ee
Note that, as it is further explained in Paper I, 
this gauge for $\g_{ab}$ can always be chosen. Here $b=b(t,y)$ is a new, 
independent scale factor.

So far the discussion was independent of the existence of branes.  
We now introduce brane
sources, which are, in general, located at specific points in the 
fifth dimension. As in Paper I, we are going to consider the setup 
motivated by heterotic M--theory. The extra dimension is compactified 
on the orbifold $S_1/Z_2$. That is, we compactify the fifth dimension
on a circle restricting the corresponding coordinate $y$ to the 
range $y\in[-R,R]$ with the endpoints identified. The action of the
$Z_2$ symmetry on the circle is taken to be $y\rightarrow -y$.
Associated to this symmetry there are two fixed points at
$y=y_1=0$ and $y=y_2=R$, where we assume that two three-branes,
stretching across $3+1$--dimensional space, are located.

We have to truncate the five-dimensional metric in order to make it 
consistent with the orbifold symmetry. For the {\it unperturbed 
metric} this implies the following conditions at the fix points:

\begin{eqnarray}
g_{\mu\nu}(-y) &=& g_{\mu\nu}(y),\label{sym1} \\
g_{\mu 5}(-y) &=& -g_{\mu 5}(y),\label{sym2}\\
g_{55}(-y) &=& g_{55}(y)\label{sym3}.
\end{eqnarray}
When the background geometry is perturbed, however, the branes are in general 
no longer located at $y_n= $constant, but at a perturbed positions
which depend on the intrinsic brane coordinates $x^i$ and $t$.
Thus, the symmetry conditions (\ref{sym1})-(\ref{sym3}) cannot 
directly be applied to the perturbed background.
However, by a suitable coordinate change, one can always gauge away 
these perturbations on the brane positions (see the discussion below).
Therefore, using these particular coordinates, 
(\ref{sym1})-(\ref{sym3}) also hold for the {\em perturbed metric}. 

Note that the conditions (\ref{sym1})-(\ref{sym3}) impose some
restrictions on the gauge transformation (\ref{gaugetransf})
that we are allowed to perform, if the new coordinate system 
is assumed to have the same symmetries. Indeed, in this case 
we have to make sure that gauge transformations do not lead 
out of the class of metrics defined this way. In particular, 
this means that the parameter $\x^\a$ for an infinitesimal 
coordinate transformation has to be restricted by
\bba
 \x^\m (-y) &=& \x^\m (y) \label{xi1}\\
 \x^5 (-y) &=& -\x^5(y) \label{xi2}\; ,
\eea
which directly follows from (\ref{gaugetransfg1})-(\ref{gaugetransfg4})
and the symmetry conditions (\ref{sym1})-(\ref{sym3}).
>From these rules we can deduce
the $Z_2$ properties of the various scalar
quantities in metric~(\ref{met5}): while the
background scale factors $a$, $b$ as well as the perturbations $\phi$,
$\psi$, $\G$, $E$ and $B_0$ are $Z_2$ even, that is, for example, $a(-y)=a(y)$,
the perturbations $W$ and $B_5$ are $Z_2$ odd, that is for example
$W(-y)=-W(y)$. Similarly,
for the scalar components in the transformation parameter $\x^\a$,
we find that $\x_0$ and $\x$ are
even while $\x_5$ is odd. 

To be precise, as we did in Paper I,
we are going to work in the {\em boundary picture}, where instead
of working with the whole orbifold, only a half of the circle is
considered. We will require that all components of metric
(\ref{met5}) are continuous across the full orbicircle except for the odd
components $W$ and $B_5$ which are allowed to jump at the fixed points
(but continuous otherwise). To simplify the notation, we define the
value of an odd field on the brane as the one that is approached from
within the interval $y\in[0,R]$. This is precisely the boundary value
of the field as viewed in the boundary picture and it represents 
{\em one half of the jump} of this field at the fixed point.

\subsection{Relocating the brane positions: a gauge for
non--bending branes}

In certain cases, the formalism of brane world perturbations is much 
simpler if the position of the branes remain fixed
at constant $y$. Indeed, if the branes do not
bend into the fifth dimension, then the $Z_2$ symmetry along the
orbifold will be straightforwardly imposed by the conditions
(\ref{sym1})--(\ref{sym3}). Furthermore, as explained in Paper I,
the low energy--effective action may easier be obtained 
if the brane positions are unperturbed.

In this subsection we will show that it is always possible to gauge 
away the perturbation of the brane position. We then motivate a novel 
gauge, in which brane world perturbations can be treated very generically. 
We will start considering a gauge for a background with two branes
located at fixed $y=y_n$ positions. Then, the most general first order
perturbation in this scenario will be described by the metric
(\ref{met5}), together with two first order perturbations on the brane 
positions. That is, the branes will bend into the fifth dimension 
according to,

\bb\label{yln}
y^{(n)}_{b}=y_n+\l_n (t, x^i).
\ee
Hence, the most general set of first order perturbations is 
$\phi$, $\psi$, $\G$, $W$, $E$, $B_a$ and $\l_n$. In that case, 
the $Z_2$--symmetry applies at the new brane positions $y^{(n)}_{b}$
and, thus, the simple conditions (\ref{sym1})--(\ref{sym3}) no longer
apply. However, this set of perturbations is too large. In fact, observe
for instance that we can construct two new gauge invariant 
quantities out of $E_{|5}$, $B_5$ and $y^{(n)}_{b}$ ($n$=1,2) by 
the following simple combination,
\bb\label{BELam}
B_5-E_{|5}-y^{(n)}_{b},
\ee
which directly follow from (\ref{gaugetransfg3})--(\ref{gaugetransfg4}) 
and the transformation rule for $y_b$: $\delta y_b = \xi_5$ after a 
first order gauge transformation (\ref{gaugetransf}). 
Thus, if we choose $B_5-E_{|5}=0$ {\it everywhere} in the bulk, 
i.e. if we choose the GLG, both brane positions are in general 
perturbed. However, as it will become clear in the following section, 
only when the anisotropic stresses on the branes vanish,
can we use the GLG and yet keep the branes at $y_b={\rm const.}$.
Hence, in the GLG the
position $y^{(n)}_{b}(x^\m)$ of the branes contain {\em only}
information about the anisotropic stresses.

Observe that, without loosing essential degrees of freedom, 
it is always possible to perform a gauge transformation that resets 
the perturbed branes to their initial unperturbed positions $y=y_n$. 
The procedure is the following:

The general metric (\ref{met5}) contains five degrees of freedom
associated to the arbitrary choice of the five quantities $\xi^\a$
that describe a five--dimensional gauge transformation. However,
in order to relocate the brane positions, we need only to partially
fix the arbitrary function $\xi^5$. 
In fact, since under a general gauge transformation 
$\x^\a$ the fifth coordinate $y$ changes according to ${\tilde y}=y+\x^5$,
we can take,

\bb
\x^5(x^\a )=\x^5_\l(x^\a )+{\hat\x}^5(x^\a ),
\ee
being $\x^5_\l(x^\a )$ {\em any odd} function across the {\em unperturbed}
brane positions, but smooth otherwise, with the following boundary
conditions at the {\em unperturbed} brane positions

\bb
\x^5_\l(t, x^i, y_n) = \mp\l_n(t,x^i),\label{x5l}
\ee
where the upper (lower) sign applies when the position $y=y_n$ is
approached to the right (left). On the other hand,
${\hat\x}^5(x^\a )$ is {\em any  smooth} function with the following
boundary conditions at the {\em unperturbed} brane positions,

\bb
{\hat \x}^5(t, x^i, y_n) = 0\label{hatx5}.
\ee
Note that the boundary condition (\ref{x5l}) allows to gauge
away the bending of the brane by fixing again the position of the
brane at $y=y_n$ (see Fig 1.), while the term ${\hat \x}^5$
with boundary (\ref{hatx5}) contains a residual gauge freedom which
still remains after fixing the position of the branes. Obviously,
any further gauge transformation should satisfy (\ref{hatx5}).
The reason is because the $Z_2$--symmetry applies again to the 
unperturbed brane positions $y_n$. Thus, $\hat\xi^5$ is an odd function
across $y=y_n$ and to avoid the appearance of distributional terms
on the metric coefficients, $\hat\xi^5$ must indeed be zero at
$y=y_n$.

While we relocate the position of the branes, we may choose two 
smooth functions $\x =E$
and $\x _0=E_{|0}-B_0$ by means of which ${\tilde E}={\tilde B}_0=0$.
Then, we have the following generic set of 
perturbations $\tilde\phi$, $\tilde\psi$,
$\tilde\G$, $\tilde W$, ${\tilde E}=0$, ${\tilde B}_0=0$ and by 
(\ref{gaugetransfg4})

\bb\label{B5xi}
{\tilde B}_5=B_5+\x_5-\x_{|5}=B_5-E_{|5}+\x_{\l\, 5}+{\hat\x}_5.
\ee
However it is not in general possible to choose a quantity
${\hat\x}_5$, that vanishes on both branes, by means of which ${\tilde
B}_5=0$. Indeed, to do so we must choose $\x _5=E_{|5}-B_5-\x_{\l\, 5}$.
However, the {\em odd} quantities $E_{|5}-B_5-\x_{\l\, 5}$ do not in general 
vanish on the branes. Therefore, the combinations
$\pm(E_{|5}^{(n)}-B_5^{(n)}+\l_n/b^2)$, 
which represent {\em half of the jump}
of the odd functions $E_{|5}-B_5-\x_{\l\, 5}$ across the branes, are already
gauge invariant. Indeed, we have not further gauge freedom to
change them. Obviously, this is a simple consequence of (\ref{BELam})
being  gauge invariant. On the other hand, as we
will explicitly see in the next section, $E_{|5}^{(n)}-B_5^{(n)}-\x_{\l\, 5}$
are directly related to the {\em anisotropic stresses} on the branes.

\begin{figure}
\hspace{3.5cm}\psfig{file=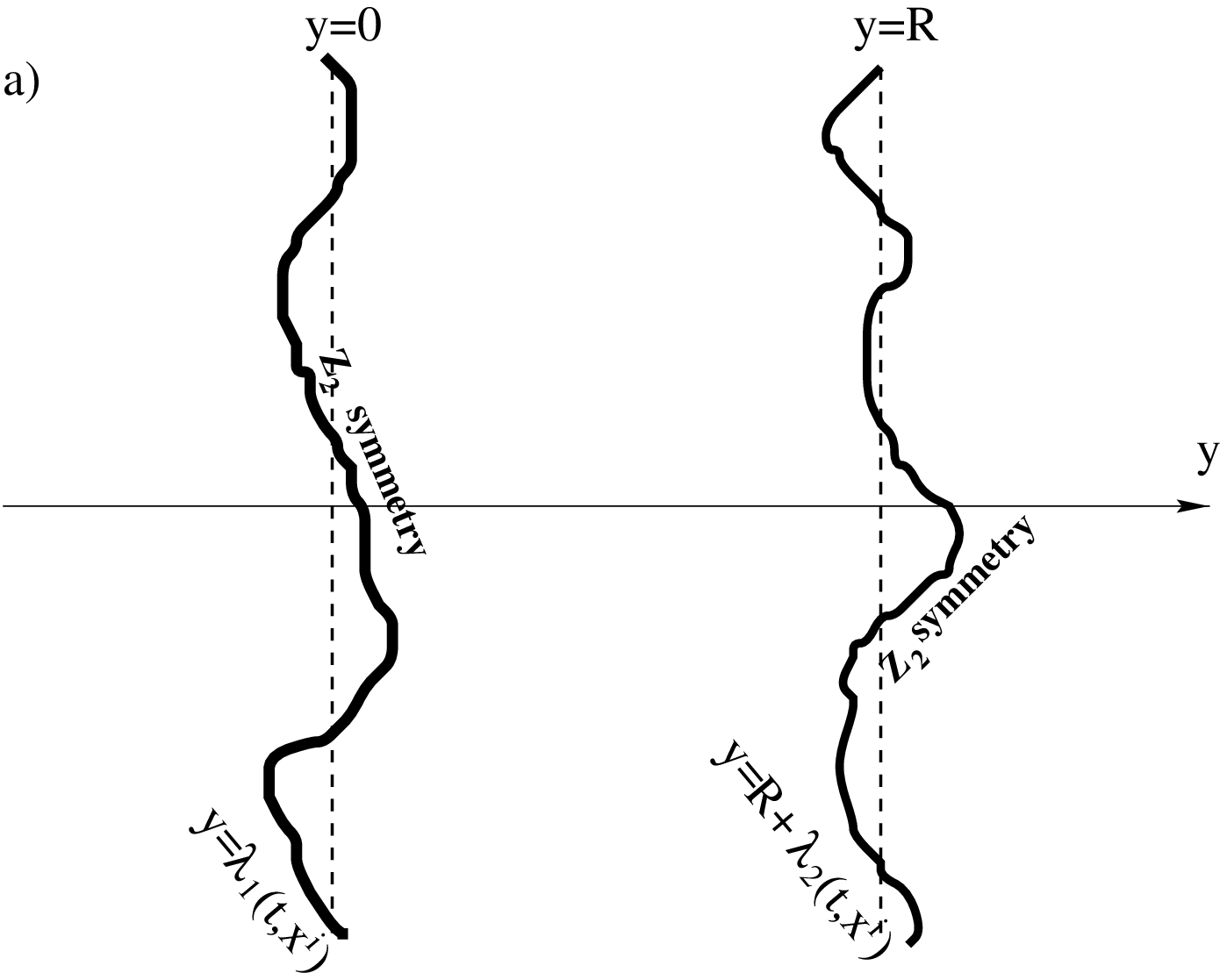,width=10cm}
\end{figure}

\begin{figure}
\hspace{3.5cm}\psfig{file=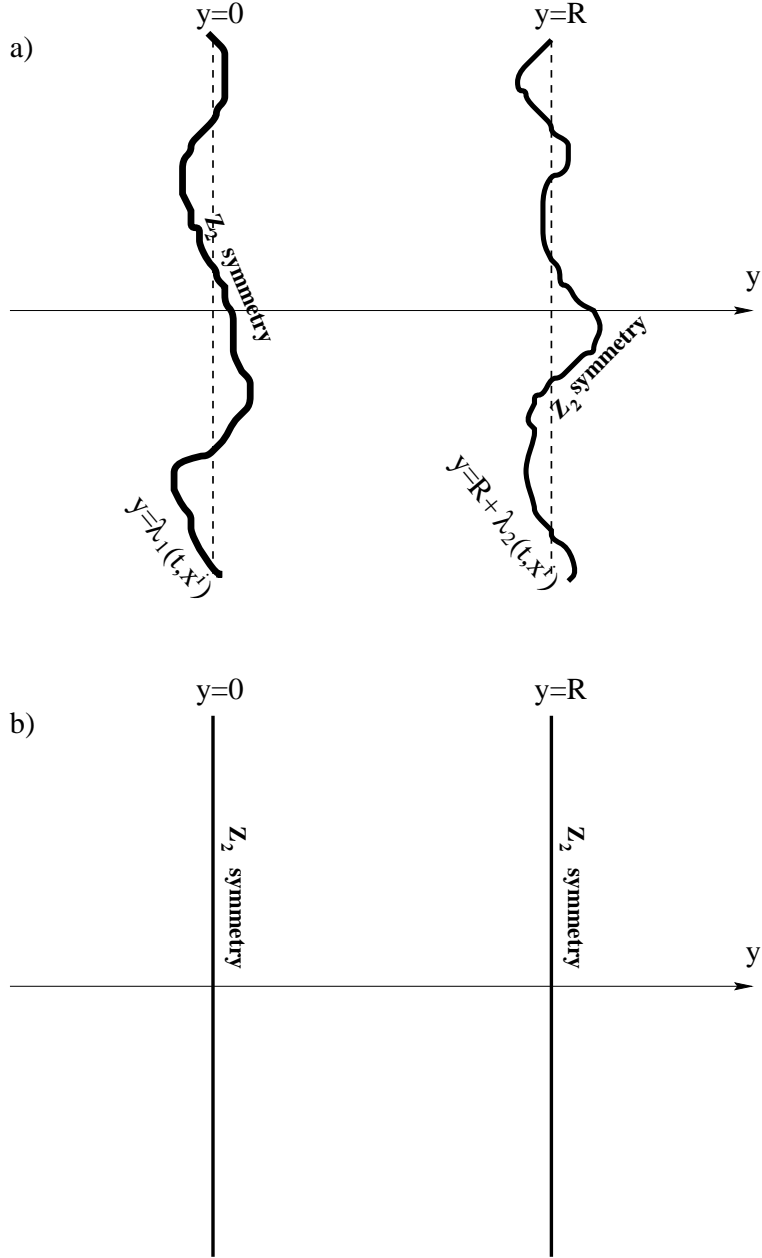,width=10cm}
\caption[h]{In the most general situation, the brane positions along 
the fifth dimension $y$ are perturbed (a). 
However without affecting the essential gauge freedom,
it is always possible 
to gauge away the perturbation on the brane positions.
Once the branes have been relocated, we can use the remaining
gauge freedom to set $E=B_0=0$ and simplify $B_5$ as much as possible.
Observe that, by relocating the brane positions at $y={\rm constant}$, we can
straightforwardly impose the $Z_2$ symmetry of the orbifold
and then the junction conditions for the metric coefficients
can be read off directly from Einstein equations.}
\end{figure}

Although it is not possible in general to switch to a gauge where the
odd perturbation ${\tilde B}_5$ is zero, we can actually eliminate most
of it as follows. Since it is always possible to choose
${\tilde B}_5$ in the form (\ref{B5xi}), we can then split

\bb\label{tildeB5}
{\tilde B}_5={\hat B}_5-{\hat E}_{|5}+L_5+{\hat\xi}_5,
\ee
where $L_5$ is the broken line that connects the values of ${\tilde B}_5$
on both branes (from within the interval $y\in [y_1,y_2]$) and shares
the same jumps of the function ${\tilde B}_5$ across the branes.
On the other hand ${\hat B}_5-{\hat E}_{|5}$
is a smooth function which vanishes on the branes and
which determines the separation between the functions ${\tilde B}_5$ and
$L_5$ (see Fig. 1). The broken line 
$L_5$  can be easily constructed as follows

\bba
L_5 &=& 
\left[{\tilde B}_5^{(1)-}+\left(\frac{{\tilde B}_5^{(2)-}-{\tilde B}_5^{(1)+}}
{y_2-y_1}\right)(y-y_1)\right]\theta(y_1-y) +
\left[{\tilde B}_5^{(1)+}+\left(\frac{{\tilde B}_5^{(2)-}-{\tilde B}_5^{(1)+}}
{y_2-y_1}\right)(y-y_1)\right]\left\{\theta(y-y_1)-\theta(y-y_2)\right\}
\nonumber\\
&+&
\left[{\tilde B}_5^{(2)+}+\left(\frac{{\tilde B}_5^{(2)-}-{\tilde B}_5^{(1)+}}
{y_2-y_1}\right)(y-y_2)\right]\theta(y-y_2)\,\nonumber
\eea
where the plus (minus) sign in ${\tilde B}_5^{(n)\pm}$ holds for the value of
${\tilde B}_5$ when the brane is approached to the right (left).
Now, explicitly using that 
${\tilde B}_5^{(n)\pm}=\pm (B_5^{(n)}-E_{|5}^{(n)}-\l_n/b^2)$, 
such a broken line can be easily simplified as follows

\bb\label{L5}
L_5=-\left(B_5^{(1)}-E_{|5}^{(1)}-\frac{\l_1}{b^2}\right)+\sum_{n=1}^2\left[
2\left(B_5^{(n)}-E_{|5}^{(n)}-\frac{\l_n}{b^2}\right)\theta(y-y_n) -
\left(B_5^{(n)}-E_{|5}^{(n)}-\frac{\l_n}{b^2}\right)\frac{y-y_1}{y_2-y_1}\right],
\ee
Observe that all the physical information that the perturbation
${\tilde B}_5$ carries is actually encoded in $L_5$. This is
true because, ${\hat B}_5-{\hat E}_{|5}$ is a smooth function that 
vanishes on the branes and thus it can be gauged away by the simple gauge 
transformation
\bb\label{hatxi5}
{\hat \x} _5={\hat E}_{|5}-{\hat B}_5
\ee
in (\ref{tildeB5}). Then, after this last gauge transformation,
the value of the perturbation ${\tilde B}_5$,
which for simplicity we will denote by $B$, would be reduced to its simplest
value, namely 

\bb
{B}=L_5.
\ee
We note here, that this particular gauge reduces to the usual 
GLG when the jumps of
the metric perturbation ${\hat B}_5$ vanish on the branes. We will see
below that this is the case when the anisotropic stresses on the
branes vanish.

With all this information in mind, we should be careful with the
construction of the complete set of gauge invariant variables
discussed in the previous section.
Of course, in order to avoid
distributional terms in the construction of the gauge invariant
variables, then (\ref{gaugeinvsca}) should be constructed 
using ${\hat B}_5 -{\hat E}_{|5}$ instead of $B_5 - E_{|5}$. Indeed,
that would prevent the appearance of delta--terms coming from 
the $y$--derivative of the odd quantity $B_5 - E_{|5}$, which does
not in general vanish on the brane positions.
  
Observe that, since ${\hat B}_5 -{\hat E}_{|5}$ is taken to be zero in
this gauge, the metric perturbations $\tilde\phi$, $\tilde\psi$, 
$\tilde\G$, $\tilde W$ 
actually correspond to the gauge invariant variables. On the other 
hand, $B$ is 
determined by the broken line (\ref{L5}), whose slope depends 
on the variables $B_5^{(n)}-E_{|5}^{(n)}$, 
which are defined on the branes only and are gauge--invariant 
under coordinate transformations which respect the symmetry 
conditions $(\ref{sym1})-(\ref{sym3})$. 

In summary, by means of a gauge transformation compatible with
leaving the branes at their unperturbed positions, 
we have been able to eliminate two 
degrees of freedom ($E$ and $B_0$) in the metric (\ref{met5})
instead of the three degrees of freedom ($E$, $B_0$ and $B_5$) that
can be cancelled with an ordinary gauge transformation. The remaining
degrees of freedom that cannot be cancelled are the jumps of
$B_5-E_{|5}+\xi_{\l 5}$ across the branes, which, as we will see in the next 
section, are directly related to the {\em anisotropic stresses} on the
branes.

For simplicity in our notation we will denote the set of gauge
invariant variables in the gauge discussed above by
$\phi$, $\psi$, $\G$, $W$ and $B$.
It is worth noticing that, since it is always possible to keep the
two branes at fixed $y=y_n$ orbifold positions,
such a gauge provide great advantages
in solving the full five--dimensional dynamics. As it will become
clear in the following section, the dynamics of the
variable $B$ is described by the evolution of the anisotropic stresses
on the branes only. Thus, when $B=0$, this particular gauge will reduce to 
the generalized longitudinal gauge introduced in Paper I.

\begin{figure}
\hspace{3.5cm}\psfig{file=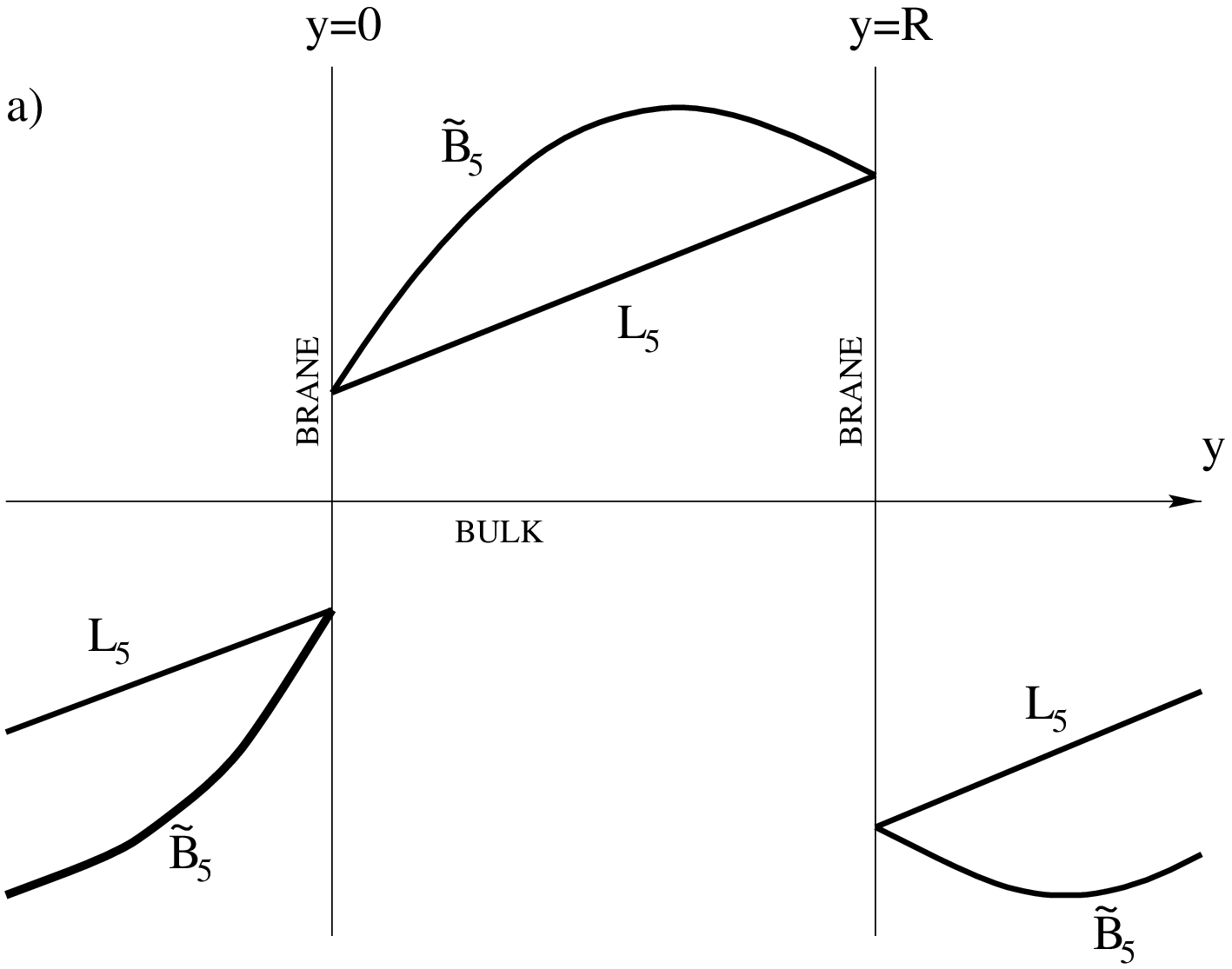,width=10cm}
\end{figure}
\begin{figure}
\hspace{3.5cm}\psfig{file=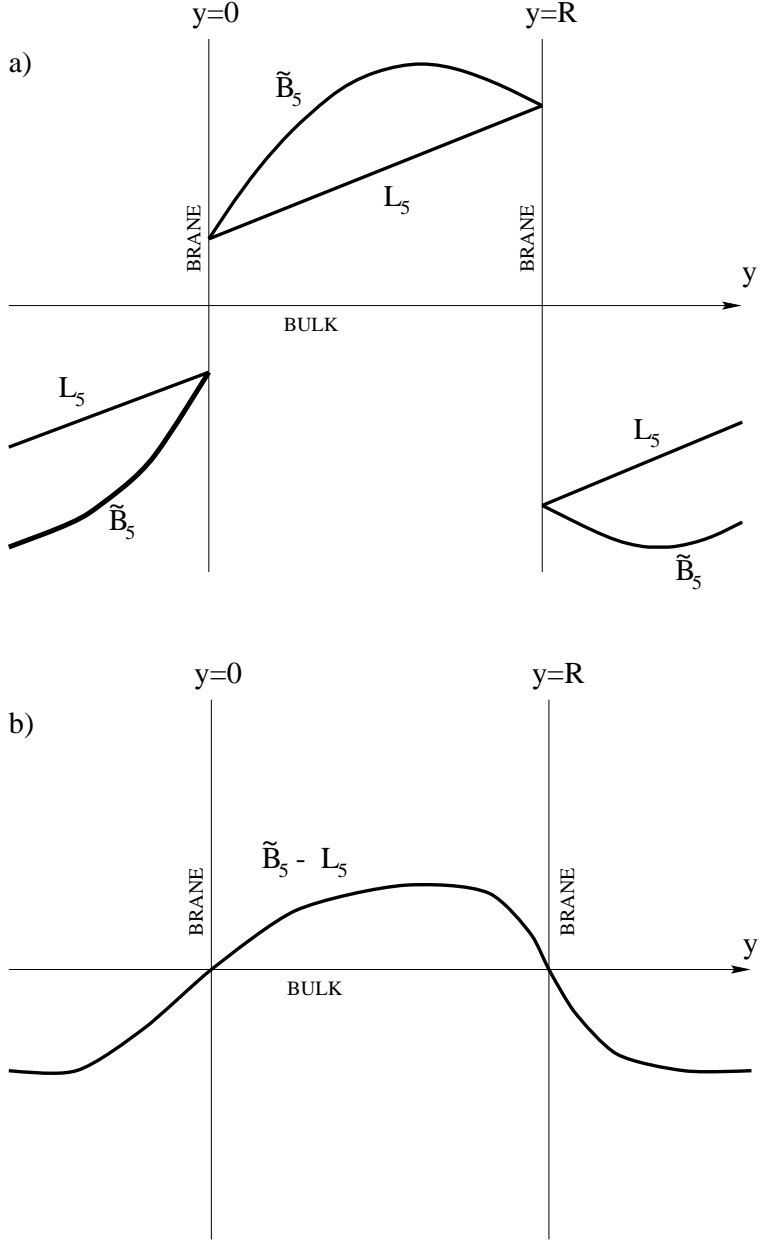,width=10cm}
\caption[h]{In general the metric perturbation ${\tilde B}_5$ cannot be gauged
away by means of a gauge transformation which is compatible with the
$Z_2$ symmetry along the orbifold. The reason is that ${\tilde B}_5$
is an {\em odd} quantity and thus it may jump on the brane positions (a).
Since the $Z_2$ symmetry conditions (\ref{sym1})--(\ref{sym3}) 
forbid the jumps of ${\tilde B}_5$ from
being gauged away, these jumps are already gauge invariant.
However, any behavior of the perturbation ${\tilde B}_5$
in the bulk  besides the purely linear is totally irrelevant.
In fact, it is always possible to perform a gauge transformation
compatible with the $Z_2$ symmetry which reduces ${\tilde B}_5$
to the straight line $L_5$ in (a). This is simply possible by choosing
a gauge transformation with ${\hat\xi}_5=-({\tilde B}_5-L_5)$. Observe that
such a ${\hat\xi}_5$ vanishes on both branes (b) and thus is compatible with
the $Z_2$ symmetry.}
\end{figure}

\subsection{Other possible gauges}

In the previous subsection we have used the gauge freedom to 
cancel the perturbations $E$ and $B_0$ and to simplify the 
perturbation $B_5$ as much as possible and yet keep the branes at 
their unperturbed positions. We have also said that this particular 
gauge choice corresponds to the generalization of the 
four--dimensional longitudinal gauge. On the other hand, another 
choice for the gauge fixation that is commonly found in 
the literature is setting the five--dimensional 
perturbations $\G$, $W$ and $B_5$ to zero, and keep the 
perturbations $E$ and $B_0$ instead, i.e. the {\em the 
scalar part of the Randall--Sundrum gauge}. However, this 
choice of gauge is generally incompatible with 
keeping the branes at their unperturbed positions. In fact, 
let us suppose that we start from the 
most general set of perturbations, i.e $\phi$, $\psi$, $\G$, 
$W$, $E$, $B_0$ and $B_5$, and we want to set $\G$ to zero. 
In order to do that, we will use the transformation equation
(\ref{gaugetransfg1}), which can be written as,

\bb\label{Gammazero}
b^2\d\G = \xi_5'-\left(\frac{a'}{a}+\frac{b'}{b}\right)\xi_5
+\left(\frac{\dot a}{a}-\frac{\dot b}{b}\right)\xi_0.
\ee
Observe that, regardless of the particular value of the
gauge fixing function $\xi_0$, if we want to set the perturbation $\G$ 
to zero we need to solve a differential equation for $\xi_5$ with
the boundary conditions $\xi^{(n)}_5=0$ on both branes. However,
this is not in general possible. As a particular example, consider
the Randall--Sundrum case where $b=1$ and $a=a(y)$, then equation (\ref{Gammazero})
would simply be,

\bb
\xi_5'-\frac{a'}{a}\xi_5-\G =0,
\ee
with general solution,

\bb\label{xi5brane1}
\xi_5(t,x^i,y)=\left[{\cal C}(t,x^i)+\int{\G(t,x^i,y) \frac{dy}{a}}\right],
\ee
where ${\cal C}(t,x^i)$ is an arbitrary integration constant. Observe,
thus, that it is always possible to fix ${\cal C}(t,x^i)$ to
set $\xi_5=0$ in one of the branes but not in the two of them at the
same time. Thus, suppose that we fix ${\cal C}(t,x^i)$ in order
to set $\xi^{(2)}=0$ on the right brane located at a constant
$y=y_2$. Then on the left brane $\xi^{(1)}\neq 0$,
which means that this brane cannot be located at a constant
$y=y_1$ position but, according to (\ref{gaugetransf}), 
its new position should necessarily be

\bb
y=y_1-\frac{1}{b^2}\xi_5(t,x^i,y_1)
\ee
where $\xi_5(t,x^i,y_1)$ is given by evaluating (\ref{xi5brane1})
at $y=y_1$. Then, any gauge transformation (\ref{gaugetransf})
from this new position has its fifth component zero in agreement
with the $Z_2$--symmetry. A similar situation would happen if
we tried to set the perturbation $W=0$.

In summary, although it is possible to use the {\em Randall--Sundrum
gauge} for brane world scenarios with a single brane located
at a fixed $y$ position, it is not in general possible in this
particular gauge to fix the positions of more than one brane.
Such an effect was already pointed out in [\ref{gar00}].

\section{The form of the perturbed Einstein equation and junction
conditions}
In this section, we calculate Einstein equations in the gauge considered in 
section 2. The metric in this gauge is given by
\begin{eqnarray}\label{metB}
ds^2 &=& a^2\left\{ b^2\left[(1+2\phi)dt^2 -2Wdtdy 
       - (1-2\Gamma) dy^2 \right]\right.\nonumber\\
     &&\qquad\left. - \left[ \Omega _{ij}(1 - 2\psi)\right] dx^i dx^j
     -2B_{|i}dydx^i\right\}.
\end{eqnarray}
Note that the only difference between such a gauge and the
generalized longitudinal gauge considered in 
Paper I, is the quantity $B=B_5$. As mentioned
in the last section, and explicitly seen  below, $B=B_5$ contains 
only information about the anisotropic stresses on the branes. 

The two branes are located in the orbifold at
$y=$ constant and the Einstein equations can be written as
\begin{equation}\label{fieldeq}
G_{\a\b}\equiv R_{\a\b} - \frac{1}{2}g_{\a\b} R = T_{\a\b} 
+ \sum_{n=1}^{2}T^{(n)}_{\a\b} \delta(y-y_n),
\end{equation}
where we have set the five-dimensional Newton constant to one, for
simplicity. The delta-functions in this equation are covariant with
respect to the fifth dimension, that is, they include a factor of
$1/\sqrt{-g_{55}}$. Furthermore, $T_{\a\b}$ is the bulk stress-energy
tensor induced by fields that propagate in the full five-dimensional
space time. The brane stress-energy tensors $T^{(n)}_{\a\b}$, on the 
other hand, originate from fields that are confined to the branes at
the orbifold fix points. We note here, that we are not considering 
Gaussian normal coordinates. 

Next, we specify the stress-energy tensors. As described in Paper I, 
they can be found using standard methods. For the bulk, the most
general form of the unperturbed stress--energy tensor is
\bb
 {T^\a}_\b = \left(\begin{array}{ccc}
             \r&0&-r\\0&-p{\d^i}_j&0\\r&0&-q
             \end{array}\right)\; ,
 \label{T}
\ee
and the background stress-energy tensor on the branes has the form
\bb
 {T^{(n)\a}}_\b = \left(\begin{array}{ccc}\r^{(n)}&0&0\\
                  0&-p^{(n)}{\d^i}_j&0\\
                  0&0&0\end{array}\right)\; .
 \label{Tn}
\ee
For the perturbed stress--energy tensor we find 
\bb
 \d {T^\a}_\b = \left(\begin{array}{ccc} \d\r&-(\r +p)b^{-2}v_{|j}&-\d r\\
         -r{B}^{|i}+ (\r +p)v^{|i}&-\d p\,{\d^i}_j+{\s^{|i}}_{|j}&-u^{|i}\\
              \d r+2r(\f +\G )-(\r +q)W&(p-q)b^{-2}B_{|j}-b^{-2}u_{|j}&
              -\d q\end{array}\right)
 \label{dT}
\ee
where $v$ and $u$ are two potentials for ``velocity'' fields and $\s$,
satisfying ${\s^{|i}}_{|i}=0$, determines the anisotropic stress.
The perturbed brane stress-energy tensors are 
\bb
 \d {T^{(n)\a}}_\b = \left(\begin{array}{ccc}
       \d\r^{(n)}&-(\r^{(n)}+p^{(n)})b^{-2}v^{(n)}_{|j}&-\d r^{(n)}\\
(\r^{(n)}+p^{(n)})v^{(n)|i}&-\d p^{(n)}{\d^i}_j+{\s^{(n)|i}}_{|j}&-u^{(n)|i}\\
\d r^{(n)}-\r^{(n)}W&p^{(n)}b^{-2}B_{|j}-b^{-2}u^{(n)}_{|j}&0\end{array}\right)\; ,
 \label{dTn}
\ee
where, unlikely in our previous work [\ref{car00}], we have explicitly
included an anisotropic stress term ${\s^{(n)|i}}_{|j}$ in the
brane stress-energy tensors and a non--vanishing component 
$\d {T^{(n)i}}_{5}$. The reason is that, as we will see below, once
we include a perturbation ${B}$ in our formalism we get new delta
terms which should be precisely matched with a brane anisotropic
stress and a brane $\d {T^{(n)^i}}_{5}$ component.
We would like to present the equations of motion based on the metric
(\ref{metB}) and on the above stress-energy tensors that follow
from the Einstein equation~(\ref{fieldeq}).
The background equations following from~(\ref{fieldeq}) have already been
given in ref.~[\ref{car00}] and thus we should refer to
this work for details. However, since the introduction of a ${B}$
perturbation will also introduce new terms on the perturbation equations 
at linear order, which are not present in our previous work [\ref{car00}], we
explicitly give them here:

\begin{eqnarray}\label{einstein00}
(ab)^2 \delta G^0_{~0} &\equiv&  
     3\left[ 2 \frac{a'b'}{ab} - 2\frac{a''}{a}  - \frac{a'}{a}
      \frac{\partial}{\partial y}- \frac{\dot a}{a} \frac{\partial}
      {\partial t} \right] \Gamma -3\left[ \frac{\dot a'}{a} +
      2\frac{a' \dot a}{a^2} +\frac{\dot a}{a} \frac{\partial}
      {\partial y}\right] W -6\left[ 2\frac{\dot a^2}{a^2} +
      \frac{\dot a \dot b}{ab}\right]\phi\nonumber \\
&&\quad + 3 \left[ 3\frac{a'}{a}\frac{\partial}{\partial y}
    - \frac{b'}{b}\frac{\partial}{\partial y}
    -3\frac{\dot a}{a}\frac{\partial}{\partial t} 
    - \frac{\dot b}{b}\frac{\partial}{\partial t}+2kb^2\right]\psi 
    +b^2\left(2 \psi + \Gamma \right)^{|i}_{~|i} + 3 \psi'' +\left[
{\partial\over\partial y}+3\frac{a'}{a}-\frac{b'}{b}
\right]{{B}}^{|i}_{~|i}\nonumber\\
&=&a^2b^2\left\{ \d\r +\sum_{n=1}^2(\d\r^{(n)}+\G\r^{(n)})
   \bar{\d}(y-y_n)\right\},
\end{eqnarray}

\begin{eqnarray}\label{einstein55}
(ab)^2 \delta G^5_{~5} &\equiv& -6\left[ 2\frac{{a'}^2}{a^2} 
      + \frac{a'b'}{ab}\right]\Gamma 
      - 3\left[ \frac{\dot{a'}}{a} + 2\frac{a' \dot{a}}{a^2} +
      \frac{a'}{a}\frac{\partial}{\partial t} \right]W
      + 3\left[ 2\frac{\dot{a}\dot{b}}{ab} - 2\frac{\ddot{a}}{a} 
      - \frac{a'}{a}\frac{\partial}{\partial y} 
      - \frac{\dot a}{a}\frac{\partial}{\partial t}\right]\phi \nonumber\\
&&\quad +3\left[ 3\frac{a'}{a}\frac{\partial}{\partial y} + 
      \frac{b'}{b}\frac{\partial}{\partial y}-3\frac{\dot a}{a}
      \frac{\partial}{\partial t}+\frac{\dot b}{b}\frac{\partial}
      {\partial t}+2kb^2\right]\psi + b^2
      \left(2\psi - \phi\right)^{|i}_{~|i} - 3\ddot \psi
+\left[\frac{b'}{b}+3\frac{a'}{a}\right]{{B}}^{|i}_{~|i}\nonumber\\
&=&-a^2b^2\d q,
\end{eqnarray}

\begin{eqnarray}\label{einstein05}
(ab)^2 \delta G^0_{~5} &\equiv&  3\left[\frac{a''}{a} - 
     2\frac{a'b'}{ab} - 2\frac{a'^2}{a^2} \right]W 
     +3\left[ 2\frac{\dot a'}{a} - 2\frac{\dot a b'}{ab} 
     -2\frac{a' \dot b}{ab} - 4 \frac{a' \dot a}{a^2} 
     +\frac{\dot a}{a}\frac{\partial}{\partial y}\right]\phi \nonumber \\
&&\quad +3\left[\frac{\partial^2}{\partial t \partial y} -
     \frac{b'}{b}\frac{\partial}{\partial t} 
     -\frac{\dot b}{b}\frac{\partial}{\partial y}\right]\psi
     - \frac{b^2}{2}{W^{|i}}_{|i}-3\frac{a'}{a}\dot\Gamma
+ \left[\frac{1}{2}\frac{\partial}{\partial t}
-\frac{\dot b}{b}\right]{{B}}^{|i}_{~|i}\nonumber\\
&=&a^2b^2\left\{ -\d r-\sum_{n=1}^2\d r^{(n)}\bar{\d}(y-y_n)\right\},
\end{eqnarray}

\begin{eqnarray}\label{einstein0i}
(ab)^2\delta G^0_{~i} &\equiv& \left\{\left[\frac{3}{2}\frac{a'}{a} 
     + \frac{b'}{b} + \frac{1}{2}\frac{\partial}{\partial y} \right] W 
     +\left[3\frac{\dot a}{a} + \frac{\dot b}{b}\right]\phi 
     +\left[ \frac{\dot b}{b}+\frac{\partial}{\partial t}\right]\Gamma
     + 2 \dot \psi  +\frac{1}{b^2}\left[\frac{1}{2}\frac{\partial}{\partial y}+
\frac{3}{2}\frac{a'}{a}-\frac{b'}{b}
\right]{\dot B}\right\}_{|i}\nonumber\\
&=& a^2\left\{ -(\r +p)v-\sum_{n=1}^2(\r^{(n)}+p^{(n)})v^{(n)}
    \bar{\d}(y-y_n)\right\}_{|i},
\end{eqnarray}

\begin{eqnarray}\label{einsteini5}
(ab)^2\delta G^5_{~i} &\equiv& \left\{\left[3\frac{a'}{a} +
      \frac{b'}{b}\right] \Gamma+\left[ \frac{b'}{b} +
      \frac{\partial}{\partial y} \right]\phi 
      +\left[ \frac{3}{2}\frac{\dot a}{a} + \frac{\dot b}{b} 
      +\frac{1}{2}\frac{\partial}{\partial t} \right] W - 2 \psi'
      +
\frac{1}{b^2}\left[2kb^2-
\frac{3}{2}\frac{\dot a}{a}\frac{\partial}{\partial t} 
+\frac{\dot b}{b}\frac{\partial}{\partial t}
-\frac{1}{2}\frac{\partial^2}{\partial t^2}
\right]{B}\right\}_{|i}\nonumber\\
&=& -a^2\left\{(q-p){B}+u+\sum_{n=1}^2\left(-p^{(n)}{B}+u^{(n)}\right)
\bar{\d}(y-y_n)\right\}_{|i},
\end{eqnarray}

\begin{eqnarray}\label{einsteinij}
(ab)^2\delta G^i_{~j} &\equiv&\left\{\left[-6\frac{a''}{a} -
       2\frac{b''}{b} + 2\frac{b'^2}{b^2}-
       3\frac{a'}{a}\frac{\partial}{\partial y}
       - 3\frac{\dot a}{a}\frac{\partial}{\partial t} 
       - \frac{\dot b}{b}\frac{\partial}{\partial t} 
       - \frac{b'}{b}\frac{\partial}{\partial y} 
       - \frac{\partial^2}{\partial t^2}\right] \Gamma\right. \nonumber \\
&&\quad +\left[ 2\frac{b' \dot b}{b^2} - 2 \frac{\dot b'}{b}
       - 6 \frac{\dot a'}{a} -3\frac{a'}{a}\frac{\partial}{\partial t}
       - 3\frac{\dot a}{a}\frac{\partial}{\partial y} 
       -\frac{\dot b}{b}\frac{\partial}{\partial y}
       -\frac{b'}{b}\frac{\partial}{\partial t}
       -\frac{\partial^2}{\partial t \partial y}\right] W \nonumber\\
&&\quad +\left[ 2\frac{\dot b^2}{b^2} - 2 \frac{\ddot b}{b}
       - 6 \frac{\ddot a}{a} -3\frac{a'}{a}\frac{\partial}{\partial y}
       - 3\frac{\dot a}{a}\frac{\partial}{\partial t}
       - \frac{\dot b}{b}\frac{\partial}{\partial t}
       - \frac{b'}{b}\frac{\partial}{\partial y} 
       - \frac{\partial^2}{\partial y^2}\right] \phi\nonumber \\
&&\quad\left. + \left[ 6 \frac{a'}{a}\frac{\partial}{\partial y}
       - 6\frac{\dot a}{a}\frac{\partial}{\partial t} 
       + 2\frac{\partial^2}{\partial y^2}
       - 2\frac{\partial^2}{\partial t^2}+2kb^2\right]\psi 
       + 2b^2(\psi -\phi+\G )^{|k}_{~|k}\right.\nonumber\\
&&\quad\left.
+\left[3\frac{a'}{a}+\frac{\partial}{\partial y}\right]{{B}}^{|k}_{~|k}
\right\}{\d^{i}}_j\nonumber\\
&&\quad -\left[3\frac{a'}{a}+\frac{\partial}{\partial y}\right]
{{B}}^{|i}_{~|j}  
-b^2(\psi -\phi +\G )^{|i}_{~|j}\nonumber\\
&=& a^2b^2\left\{ -\d p\,{\d^i}_j+{\s^{|i}}_{|j}
    -\sum_{n=1}^2\left[(\d p^{(n)}+\G p^{(n)})\,
    {\d^i}_j-{{\s^{(n)}}^{|i}}_{|j}\right]\,\bar{\d}(y-y_n)\right\}.
\end{eqnarray}
Here we have defined the delta-function $\bar{\d}$ which incorporates
a factor $(-g_{55})^{-1/2}=1/ab$. 

We write down also the traceless part of eq.~(\ref{einsteinij}):

\bb
(\psi -\phi +\G )^{|i}_{~|j}-\frac{1}{3}(\psi -\phi +\G )^{|k}_{~|k}
  {\d^i}_j = -a^2{\s^{|i}}_{|j}-
\frac{1}{b^2}\left(3\frac{a'}{a}+\frac{\partial}{\partial y}\right)
\left[ B^{|i}_{~|j}-\frac{1}{3}\d ^i_j B^{|k}_{~|k}
\right].
\ee
Observe that, for the particular case where the anisotropic stress
in the bulk vanishes, i.e. ${\s^{|i}}_{|j}=0$, we may identify

\bb
\psi -\phi
+\G=-\frac{1}{b^2}\left(3\frac{a'}{a}+\frac{\partial}{\partial y}\right)B.
\ee

\medskip
As a next step we analyze the background and the perturbed Einstein 
equations:
\begin{center}
\underline{{\bf Background equations }}
\end{center}
Matching first the delta terms in the background equations we can
easily get the junction conditions for the scale factors 
$a(t,y)$ and $b(t,y)$ (see Paper I for details):

\bb\label{[a,b]}
{a'\over a}=\mp{1\over 6}ab\r^{(n)}\; ,\qquad
{b'\over b}=\pm\frac{1}{2}ab\left(\rho^{(n)}+p^{(n)}\right)\; .
\ee
Here and in the following the upper sign holds for the brane 
at $y=y_1$ and the lower sign for the brane at $y=y_2$. 

On the other hand, from the 05 and 55 component of the background
equations, we find
\bba
\dot{\r}^{(n)} &=& -3{{\dot a}\over a}\left(\r^{(n)} +
                  p^{(n)}\right)\mp 2abr\; ,\label{conservB}\\
\frac{\ddot{a}}{a}-\frac{\dot{a}\dot{b}}{ab}+kb^2
  &=& -{a^2b^2\over 3}\left[\frac{1}{12}\r^{(n)}\left(\r^{(n)}+3p^{(n)}\right)
    +q\right]\; ,
\eea
which respectively represent the background
{\em energy conservation equation} on the brane and a background
dynamical equation for the scale factors on the brane.

\medskip
\begin{center}
\underline{{\bf Perturbed equations }}
\end{center}
Observe first that the components
$\delta G^0_{~0}$, $\delta G^i_{~i}$, $\delta G^i_{~j}\; (i\neq j)$, 
$\delta G^0_{~i}$, $\delta G^0_{~5}$ and $\delta G^5_{~i}$ contain explicit
delta-function terms and they should be matched by terms containing first
$y$ derivatives of $W$ and ${B}$ and second $y$
derivatives of all other quantities. This leads to
\bba
 \psi ' &=& {{\dot a}\over a}W-
          \frac{1}{3}\left(B_5^{(n)}-E_{|5}^{(n)}\right)^{|k}_{~|k}
          \pm {1\over 6}ab\left(\delta\r^{(n)}
            -\Gamma\rho^{(n)}\right), \label{jump1}\\
 \phi ' &=& -\left({{\dot a}\over a}+{{\dot b}\over b}+{\partial\over
            \partial t}\right)W\pm {1\over 3}ab\left(\delta\r^{(n)}
            -\Gamma\r^{(n)}\right)\pm\frac{1}{2}
            ab\left(\delta p^{(n)} - \Gamma
              p^{(n)}\right),\label{jump2}\\
\pm\frac{ab}{2}\,{\s^{(n)|i}}_{|j} &=& 
            -\left(B_5^{(n)}-E_{|5}^{(n)}\right)^{|i}_{~|j}
         +\frac{1}{3} {\d^i}_j
         \left(B_5^{(n)}-E_{|5}^{(n)}\right)^{|k}_{~|k},
       \label{jump22}\\
 W &=& -\frac{1}{b^2}\left({\dot B}^{(n)}_5-{\dot E}^{(n)}_{|5}\right)
       \mp\frac{a}{b}\left(\r^{(n)}+p^{(n)}\right)
        v^{(n)},\label{jump3} \\
 W &=& \frac{\d r^{(n)}}{\r^{(n)}}, \label{jump4} \\
B_{|i}^{(n)} &=&\left(B_5^{(n)}-E_{|5}^{(n)}\right)_{|i}=
                 \frac{u^{(n)}_{|i}}{p^{(n)}}.\label{jump5}
\eea
In these equations the background junction conditions (\ref{[a,b]}) have 
been used. From eqns. (\ref{jump3}) and (\ref{jump4}) we find 

\bb\label{drn}
 \d r^{(n)} =
   \mp\frac{a}{b}\r^{(n)}\left(\r^{(n)}+p^{(n)}\right)v^{(n)}
-\frac{\r ^{(n)}}{b^2}\left({\dot B}^{(n)}_5-{\dot E}^{(n)}_{|5}\right)\; .
\ee
Recall that the quantities $\d
 r^{(n)}$ and $u^{(n)}$ are uniquely fixed by the other components and they
are, generally, non-vanishing, as already discussed in Paper I.
In fact, they
are directly related to the anisotropic
stresses on the branes. For instance, we can easily transform
(\ref{drn}) into an equation involving the anisotropic stresses on the
brane as follows:

\bb\label{drns}
{\d\Pi^{(n)}}^i_{~j}=
\pm \frac{1}{2}\frac{\r ^{(n)}}{b^2}
\frac{\partial}{\partial t}\left(ab{\s ^{(n)}}^{|i}_{~|j}\right)
\mp \frac{a}{b}\r^{(n)}\left(\r^{(n)}+p^{(n)}\right){V^{(n)}}^i_{~j},
\ee
where we have introduced for convenience the traceless quantities
${\d\Pi^{(n)}}^i_{~j}$ and ${V^{(n)}}^i_{~j}$, which are given by

\bba
{\d\Pi^{(n)}}^i_{~j} &=&
{\d r^{(n)}}^{|i}_{~|j}-\frac{\d ^i_j}{3}{\d r^{(n)}}^{|k}_{~|k},
\label{drnpi}\\
{V^{(n)}}^i_{~j} &=&
{v^{(n)}}^{|i}_{~|j}-\frac{\d ^i_j}{3}{v^{(n)}}^{|k}_{~|k}.
\eea

Finally, we would like to discuss the influence of the new terms 
discussed in this section (i.e. the quantities $B=B_5$ and
$\sigma^{(n)}$) for the low--energy effective theory discussed in 
Paper I. Because $B$ in the bulk simply connects the values of the 
anisotropic stresses on the branes (see the line--construction in 
section 2), the $y-$averaged $B$ along the bulk is zero. 
However, at the position 
on the branes, the jumps of the function have to be taken into account, 
so that the effective four--dimensional anisotropic stress $\sigma_4$ 
has to be replaced by 

\begin{equation}
\sigma_4 = e^{-\chi}\left< \sigma \right> + \frac{1}{2R e^{2\chi}} 
\sum_{n=1}^{2} \sigma^{(n)},
\end{equation}
where $<...>$ describes the averaging over the fifth direction. 
The other terms are not affected ($R$ is the coordinate distance between the 
branes and $\chi$ is a modulus describing the size of the fifth 
dimension).

\section{Gravitational interaction between the branes and the bulk}

In this section, we would like to discuss the gravitational 
interaction between the branes and the bulk. Since our gauge is well
adapted to analyze the whole five--dimensional brane world, we  
should be able to get some insights on the non-local nature
of this particular geometry.

Observe, that the variables $\d r^{(n)}$ are the
components ${\d T^{(n)0}}_{5}$ of the stress--energy tensor
(\ref{dTn}) on the
branes. Thus, $\d T^{(n)05}=\d r^{(n)}/(a^2b^2)$ is connected to 
an energy flux, i.e. it determines a flux of energy along the fifth 
dimension. Observe from (\ref{drnpi}) that the anisotropic part of
$\d r^{(n)}$ is directly related to quantities like
$(\r^{(n)}+p^{(n)})V^{(n)}$ or 
${{\dot \s}^{(n)}}$, which are naturally related
to the  third order time--derivative of the {\em reduced quadrupole
moment} of the matter fields on
the brane\footnote{The reduced quadrupole moment of the matter fields on
the brane is defined as,
${\barI}^{ij}=\int{\r\left[x^ix^j
-\frac{1}{3}\d^{ij}x^kx_k\right]d^3x}$.
Thus, ${\ddot {\barI} }^{ij}$ is related to the 
{\em non--spherical part of
the kinetic energy of the source} which is related to the anisotropic
stress.}. Observe, that the emission of gravitational
radiation by a  cosmological source is directly linked to the third
order time--derivative of its reduced quadrupole moment 
(see [\ref{mis}] for details).
Therefore, we may physically interpret the anisotropic part of
the quantities  $\d r^{(n)}$ as a non--local measure of the energy
exchange between the branes and the bulk through the exchange of
five--dimensional gravitational radiation. 

The generalization of the energy--momentum conservation equation
(\ref{conservB}) for brane perturbations
can be found by restricting the $05$ and $5i$--components of 
the perturbed Einstein equations to the branes. This respectively
gives the {\em continuity equation} and {\em Euler's equation}.
These two equations together with (\ref{drn}) give the full
energy--momentum conservation for the brane matter field perturbations
under possible interactions with the bulk. In fact,
making explicit use of the junctions equations (\ref{jump3})--(\ref{jump5})
these three equations may be conveniently written as follows

\bb\label{econ0}
\frac{\partial}{\partial t} \delta \rho^{(n)} + 
3 \frac{\dot a}{a} (\delta \rho^{(n)}
+ \delta p^{(n)} ) - 3 \frac{\dot a}{a} (\rho^{(n)} + p^{(n)})\dot \psi =
- (\rho^{(n)} + p^{(n)})v^{|k}_{~|k}  \pm 2 ab (\rho + q) 
\delta r^{(n)} + 2 ab (\delta r + 2\phi r + \Gamma r),
\ee

\bba\label{euler}
& &\frac{\partial}{\partial
  t}\left[\left(\rho^{(n)} + p^{(n)} \right)v^{(n)}_{|i}\right] + 
2\left(2\frac{\dot a}{a}-\frac{\dot b}{b}\right)
\left(\rho^{(n)}+p^{(n)}\right)v^{(n)}_{|i} =\\
 & & - b^2\left[\left(\rho^{(n)}+p^{(n)}\right)\f_{|i} +\d p^{(n)}_{|i}
\right]
 \mp\frac{4}{ab}\left[a^2 b^2 (p-q)B^{(n)}_{|i}+b^2kB^{(n)}_{|i}+\frac{1}{3}
b^2 {B^{(n)|k}}_{|ik} + \frac{1}{2}a^2 b^2 u_{|i}\right],\nonumber
\eea 

\bb\label{rnB}
\frac{\partial}{\partial t}B^{(n)}=\mp ab
\left(\rho^{(n)}+p^{(n)}\right)v^{(n)}-b^2\frac{\d r^{(n)}}{\r ^{(n)}}
\ee
These equations provide a coupled system of differential
equations for the matter perturbations on the brane.
In particular observe that:

\begin{itemize}

\item[i)] Equation (\ref{econ0}) describes the growth of pressure
and density perturbations sourced by $r$, $\d r$, $\d r^{(n)}$
and a Poisson--like term $v^{|k}_{~|k}$ from the velocity
perturbations.

\item[ii)] Equation (\ref{euler}) describes the growth of velocity
perturbations, which are sourced by $\d p^{(n)}$, $\f$, the anisotropic
stresses on the branes and the momentum flux $u_{|i}$ 
from the bulk matter. 

\item[iii)] Equation (\ref{rnB}) describes the growth of anisotropic
stresses on the branes sourced by $\d r^{(n)}$ and velocity
perturbations on the branes. This equation states that anisotropic
stresses can be smoothed out by changing into velocity perturbations
and gravitational radiation.
For the particular case that $\d r^{(n)}=0$, then equation (\ref{drn})
states that decays on the anisotropic stresses can induce velocity  
fluctuations on the matter fields only (reference [\ref{whu98}] 
provides details on this type of effects for ordinary cosmology). 

\end{itemize}

Let us collect the physical meaning of the
quantities $r$, $\d r$ and $\d r^{(n)}$:

\begin{itemize}
\item[i)] $r$ determines a non-perturbed energy flow in the bulk
  arising from the
$05$--component of the background stress--energy tensor (\ref{T}).
For the particular case that the bulk is empty, i.e. $\r+q=0$, this
term is zero.

\item[ii)] $\d r$ determines a first order perturbation on the energy
flow in the bulk arising from the $05$--component of the perturbed 
stress--energy tensor (\ref{dT}). Also, for the particular case that the
bulk is empty this term is zero. Recall, however, that being $\d r=0$
does not mean that there is not a flow of energy through the bulk.
In fact, it is possible to have {\em pure gravitational radiation}
propagating in the bulk and yet have $\d r=0$. This is true because
$\d r$ comes from a purely {\em local} stress--energy tensor for
first order bulk perturbations. However, the energy
associated to gravitational waves is {\em non--local}. Indeed,
first order gravitational waves are vacuum solutions of Einstein
equations and their energy can be only properly computed by non--local 
space--time averages of quadratic wave amplitudes
(see for instance [\ref{mis}])\footnote{
The propagation of linear gravitational waves is determined by
the first order perturbation of the {\em vacuum} Einstein equations.
However, the {\em backreaction} of the gravitational waves over the 
background geometry is determined by Einstein equations with a non--zero
stress--energy tensor associated to the waves. Such a tensor can be
obtained by non--local averages of quadratic combinations of the first
order gravitational wave amplitudes.}.

\item[iii)] $\d r^{(n)}$ determines a first order perturbation of the
``projected'' energy flow on the branes, which may have two
contributions: a local one (due to the matter distribution) and 
a non--local one (due to gravitational waves). While the local 
projection of $\d r^{(n)}$ on the brane is zero when the bulk is empty, 
in general $\d r^{(n)}\neq 0$ due to the non--local contributions.
It is worth noticing that for domain walls we have $\delta r^{n} = 0$, 
because the equation of state is $\rho+p=0$ and the anisotropic 
stress vanishes. This suggest, that, at least for the scalar part 
discussed here, domain walls do not radiate energy (at first order), 
see also [\ref{domain}].

\end{itemize}

Finally, the restriction of the $55$ component of the perturbed Einstein 
equations, eq.~(\ref{einstein55}), to the brane leads to the 
following evolution equation
\bba
 b^2(2\psi -\f )^{|i}_{~|i}&-&3\ddot{\psi}-3\frac{\dot{a}}{a}\dot{\f}
 +3\left(\frac{\dot{b}}{b}-3\frac{\dot{a}}{a}\right)\dot{\psi}
 +6kb^2(\psi+\phi)+a^2b^2{\r^{(n)}}^2\left[\frac{1}{6}(1+3w^{(n)})\f
 \right.\nonumber\\
 &+&\left.\frac{\d q+2q\f}{{\r^{(n)}}^2}+\frac{1}{6}\left(
 1+\frac{3}{2}w^{(n)}\right)\d^{(n)}+\frac{\d p^{(n)}}{4\d\r^{(n)}}
 \d^{(n)}\pm\frac{a}{b}\frac{r}{\r^{(n)}}(1+w^{(n)})v^{(n)}\right]
+a^2r\left({\dot B}^{(n)}_5-{\dot E}^{(n)}_{|5}\right) = 0\; .
\eea

\section{Discussion} 

Using the formalism presented in Paper I, we have discussed how
anisotropic stresses on the branes can be easily included. 
We have also argued, that if we use the generalized longitudinal
gauge in the bulk, the branes no longer remain 
at fixed $y={\rm const.}$ orbifold positions. On the other hand,
the brane positions are related to the anisotropic stresses only.
Also, we have introduced a novel gauge, in which the positions of the
branes may be fixed at $y={\rm const.}$ while the information on
the anisotropic stresses is encoded into the metric perturbation
$B$. Finally, we have discussed the gravitational interaction between 
the branes, which, in our formalism, is described by the 05--component 
of the brane energy momentum tensor. 

An outstanding problem is 
to find solutions of the bulk Einstein equations 
(\ref{einstein00})--(\ref{einsteinij}).
It is worth noticing that some work has already addressed this issue
from the point of view of a brane--observer only. However, in the
presence of perturbations, this is not enough. In fact, one needs to
solve the full five--dimensional Einstein equations in order to
fully understand the non--local interaction between the two branes.
Some work along this direction was already done in Paper I for the 
linear M--theory background. 
A derivation of the low-effective perturbation 
equations in Randall--Sundrum brane world [\ref{RS}] (with two
branes) is still lacking. However, even without specifying the details
of the theory, one is able to derive rather general statements about 
the evolution of perturbations and the consequences of the dynamics 
of the perturbations,  (see [\ref{wands}], [\ref{per2}], [\ref{per4}, 
[\ref{car01}], [\ref{gordon}] and [\ref{per6}] 
for a discussion on scalar perturbations, [\ref{per5b}] for a 
discussion on vector perturbations and
[\ref{langlois}] for tensor perturbations). For example, in 
[\ref{car01}] it was shown that due to the energy--flow, onto or away from the 
branes, will modify the evolution of super-horizon amplitudes 
on the brane. In fact, in contrast to the usual results in four
dimensions [\ref{Brandenberger}], adiabatic perturbations will not be constant 
in general. As discussed in [\ref{car01}], this effect is 
a consequence of the statement that super-horizon amplitudes are 
constant if the energy-conservation holds [\ref{wands2}]. However,
energy--conservation can easily be violated in brane worlds.

Large scale anisotropies for a brane in an Anti-de Sitter bulk were 
discussed in [\ref{per6}]. Also, using the gauge
presented in this paper, the full set of 
equations on the brane for this scenario was derived in [\ref{car11}]. 
Hence, there is the hope, that the formalism
presented here can be used to fully understand the interaction between
the bulk gravitational field and the brane matter.

{\it Note added:} While this paper was prepared for publication,
ref. [\ref{per5a}] appeared, which addressed some of the questions 
in section 2. 

\begin{center}
{\bf Acknowledgments:}\\
\end{center}
We are grateful to R. Brandenberger, S. Davidson, A. Davis, N. Deruelle,
F. Finelli, D. Langlois, R. Maartens, D. Wands and J. Weller for useful 
discussions about anisotropic stresses on the branes and the 
$Z_2$--symmetry. C.~v.~d. Bruck is supported by the Deutsche 
Forschungsgemeinschaft (DFG). Miquel Dorca is supported by the 
{\em Fundaci\'on Ram\'on Areces}. The research was supported in 
part (at Brown) by the U.S. Department of Energy under Contract 
DE-FG02-91ER40688, TASK A. 

\references
\item \label{wands} R. Maartens, D. Wands, B.A. Bassett, I. Heard, Phys.Rev.D 
{\bf 62}, 041301 (2000) 
\item \label{hawking} S.W. Hawking, T. Hertog, H. Reall, Phys.Rev.D {\bf 62},
043501 (2000) and hep-th/0010232
\item \label{soda2} S. Kobayashi, K. Koyama, J. Soda, hep-th/0009160 
\item \label{car00}  C. van de Bruck, M. Dorca, R. H. Brandenberger and
A. Lukas, Phys. Rev. D. {\bf 62}, 123515 (2000) (PAPER I)
\item \label{per11} S. Mukohyama, Phys. Rev. D {\bf 62}, 084015 (2000)
and hep-th/0006146 
\item \label{per1} H. Kodama, A. Ishibashi, O. Seto, 
Phys. Rev. {\bf D62}, 064022 (2000)
\item \label{per2} R. Maartens, Phys. Rev. D {\bf 62}, 084023 (2000)
\item \label{per3} D. Langlois, Phys. Rev. D {\bf 62}, 126012 (2000)
and hep-th/0010063 
\item \label{per4} K. Koyama, J. Soda, Phys. Rev. D. {\bf 62}, 123502 (2000)
\item \label{per5} N. Deruelle, T. Dolezel, J. Katz, hep-th/0010215
\item \label{per5a} A. Neronov, I. Sachs, hep-th/0011254
\item \label{per5b} H. Bridgman, K. Malik, D. Wands, hep-th/0010133
\item \label{Brandenberger} V.F. Mukhanov, H.A. Feldman, R.H. Brandenberger, 
Phys. Rep. {\bf 215}, 203 (1992)
\item \label{car01} C. van de Bruck, M. Dorca, C.J.A.P. Martins,
M. Parry, to appear in Phys. Lett. B, hep-th/0009056
\item \label{gordon} C. Gordon, R. Maartens, hep-th/0009010
\item \label{gar00} G. Garriga and T. Tanaka, Phys. Rev. Lett. 
{\bf 84}, 2778, (2000) 
\item \label{isr66} W. Israel, Nuovo Cim. {\bf 44B}, 1 (1966)
\item \label{langlois} D. Langlois, R. Maartens, D. Wands,
Phys. Lett. B {\bf 489}, 259 (2000)
\item \label{per6} D. Langlois, R. Maartens, M. Sasaki, D. Wands, 
hep-th/0012044 
\item \label{car11} C. van de Bruck, M. Dorca, hep-th/0012073 
\item \label{wands2} D. Wands, K.A. Malik, D. Lyth, A.R. Liddle, 
Phys.Rev. D{\bf 62}, 043527 (2000) 
\item \label{whu98} W. Hu, ApJ {\bf 506}, 485 (1999)
\item \label{domain} H. Kodama, H. Ishihara, Y. Fujiwara, 
Phys. Rev. D {\bf 50}, 7292 (1994)
\item \label{bert} C.-P. Ma, E. Bertschinger, Astrophys. 
Journ. {\bf 455}, 7 (1995)
\item \label{mis} C. W. Misner, K. S. Thorne, J. A. Wheeler,
{\em Gravitation}, W. H. Freeman and Company, San Francisco (1973)
\item \label{RS} L. Randall, R. Sundrum, 
Phys.Rev.Lett. {\bf 83}, 3370 (1999) 
\end{document}